\documentclass [12pt] {article}
\usepackage {amssymb, amsfonts}
\usepackage [T2A] {fontenc}

\usepackage [cp866] {inputenc}
\inputencoding {cp866}
\textwidth = 16cm \newcommand {\bk} +   \newcommand {\ox} {e ^ {-\frac
{\omega _ 0x ^ 2} {4}}}  \newcommand {\px} {\sqrt {\frac {\omega _ 0} {2}} x}

\begin {document}


\centerline {\bf {Darboux Transformation}} \centerline {\bf for
the non-stationary Shr\"odinger Equation} \vskip0.5cm

\centerline {\bf E.Sh.Gutshabash} \vskip0.3cm

\vskip0.3cm \centerline{\small Research Institute of Physics, St.
Petersburg State University ,} \centerline{\small St. Petersburg,
Staryi Peterghof, 198504, Russia} \centerline{\small
 e-mail:
gutshab@EG2097.spb.edu} \vskip0.3cm

{\small The Lax representation for the nonstationary Schr\"odinger
equation with rather arbitrary potential is proposed. Some
examples of the construction of exact solutions are given by means
of Darboux Transformation method.}

\ \vskip0.5cm \centerline {{1}. {\bf Introduction.}} \vskip0.3cm
The increase of interest to the non-stationary Shr\"odinger
equation (NtSH) noticed during last decades is apparently
explained not only by needs to solve quantum-mechanical problems,
but also by its role in the method of inverse problem (for
example, in the theory of the Kadomtsev-Petviashvili equation) and
related problems [1-4].

It is necessary to note, that known methods of construction of the
exact solutions of NtSH equation are basically concentrated around
two methods: the method of intertwining relations (its special
case is a method of Darboux Transformation (DT)) [1,2] and the
method of dressing chains (see, for example, [3]). These methods
have both some advantages and definite defects (without details,
we shall specify only, that they do not give a regular procedure
of construction of the solutions neither for any rather "good"
\enskip potentials, nor for the case of higher spatial
dimensionality).

It is necessary to mention separately the paper [5], in which the
scattering theory was developed directly for the NtSH equation,
and its connection with the solutions of the
Kadomtsev-Petviashvili equation was discussed.

In a certain sense the alternative approach was proposed in [6].
It is based on the application of the Darboux-like anzatz, it does
not depend on the spatial derivatives (and, hence, on the
dimensionality of the problem), and it also reduces the solution
of NtSH equation to the solution of some functional-differential
equation, which does not contain any potential.

The idea of the present paper is the following. The NtSH equation
(though it is linear, but it includes a coefficient function -
potential) is considered as a (rather special) particular case of
the nonlinear Shrodinger equation:

$$ i\Psi _ t = -\Psi _ {xx} + \varkappa (\Psi, \:\bar \Psi) \Psi, \eqno (1.1)
$$
assuming that $ \varkappa (\Psi, \:\bar \Psi) \equiv W (x, \:t) $,
where $ W (x, t) $ is a given function depending only on spatial
and time arguments. Then this equation represents the completely
integrable system. It means, that it possesses also all
appropriate attributes of such systems: the scattering problem, an
infinite and involutive  series "of the conservation laws" \enskip
(since the initial system is not conservative, no sense to speak
about "the relations of balance"), identities of traces, hierarchy
of the higher equations etc. Certainly, besides that it also
allows to construct the exact solutions of the initial equation,
for which only limited class of solutions is known up to now (see,
for example, monography [7]; let's also note here the important
review [8], in which the quantum-mechanical systems were discussed
from the point of view of the representation theory of groups and
algebras).

Thus, the problem is reduced to the construction of a Lax pair and
to the analysis of the appropriate associate linear system.

The paper is organized as follows. In Section 2  we obtain the Lax
representation by means of transition to the Heisenberg picture.
Section 3 is devoted to verification of the covariance property
and to calculation of the exact solutions by the DT method.

\vskip0.5cm

\centerline {2. \bf Lax Representation and Heisenberg Picture.}

\vskip0.5cm

Let us consider the NtSH equation (here and below we use the
system of units, in which the Plank's constant $ \hbar = 1,$ and
the mass of particle$ \:m = 1/2 $) in the coordinate
representation:

$$
i\Psi _ t = -\Psi _ {xx} + W\Psi, \eqno (2.1)
$$
where $ \Psi = \Psi (x, t) $ is the wave function {\footnote {As
was noticed in [5], the replacement $ \Psi \to \Psi \exp {(\lambda
t/i)} $ reduces the equation $ i\Psi _ t = -\Psi _ {xx} + (W (x,
t) -\lambda) \Psi $, where $ \lambda \in {\mathbb C} $, to the
form in (2.1), and thus the analysis given below is applicable
also to this equation.}. Potential $ W = W (x, t) $ is supposed to
be given (it plays a role of functional parameter) and real,
therefore:

$$
i(| \Psi | ^ 2) _ t = ({\bar \Psi} _ x\Psi-\Psi _ x {\bar \Psi}) _
x. \eqno (2.2)
$$
Let's rewrite the equation (2.1) as a condition of compatibility
of linear system of ($ 2\times 2 $) matrix equations:

$$
\Phi _ x = U\Phi\Lambda, \:\:\:\Phi _ t = V\Phi\Lambda, \eqno
(2.3)
$$
where $ \Phi = \Phi (x, t, \lambda) \in Mat (2, \mathbb {C}), \:
\Lambda = diag (\lambda, \: {\bar \lambda}), \: \lambda \in
\mathbb {C} $ is the parameter, $ {\mathrm Im} \lambda \ne 0 $,
and let's assume that

$$
U = U (x, t) = \left (\begin {array} {cc} u_0 & f _ 0 \\
g _ 0 & {\bar u} _ 0 \end {array} \right), \:\:\:V = V (x, t)
=\left (\begin {array} {cc} v_0 &
 p_0 \\
q _ 0 & {\bar v} _ 0
\end {array} \right) .\eqno (2.4)
$$
Here $ u_ 0 = u _ 0 (x, t), \:v _ 0 = v _ 0 (x, t), \:f _ 0 = f _
0 (x, t), \:g _ 0 = g _ 0 (x, t), \:p _ 0 = p _ 0 (x, t), \:q _ 0
= q _ 0 (x, t) $, and the equation (2.1) is equivalent to the
relation:

$$
u_{0t} = v_{0x}. \eqno (2.5)
$$
It can be considered as the constraint, which can be solved as:

$$ u_0 = i\Psi (x, t) -\frac {\alpha} {2} \int _ 0 ^ t
W (x, t ^ {\prime}) \Psi (x, t ^ {\prime}) dt ^ {\prime} + A _ 0
(x), $$
$$
\eqno (2.6) $$
$$ v_0 = -\Psi _ x (x, t) + \frac {2-\alpha} {2} \int _ {-M} ^ x W (x ^ {\prime},
\:t) \Psi (x ^ {\prime}, \:t) dx ^ {\prime} + B _ 0 (t),
$$
where $ \alpha \in {\mathbb R}$ is the parameter, $A _ 0(x), \:B _
0 (t) $ are arbitrary functions, and $M
> 0 $ is arbitrary (finite or infinite) number, introduced in order
to provide the convergence of integrals below.

The requirement of compatibility of system (2.3) leads to
relations: $ U_t - V_x = 0 $ and $ [U, V] = 0 $; by the
straightforward calculations one can check that they will be
fulfilled, if the functions $ f _ 0, \:g _ 0 $ will satisfy the
linear equations in partial derivatives of the first order:

$$
( \ln f _ 0) _ t-\frac {1} {w _ 0} (\ln f _ 0) _ x = -\frac {1} {w
_ 0} (\ln w _ 0) _ x, \:\:\:\: (\ln g _ 0) _ t-\frac {1} {w _ 0}
(\ln g _ 0) _ x = -\frac {1} {w _ 0} (\ln w _ 0) _ x, \eqno (2.7)
$$
where $ w _ 0 = (u _ 0 - {{\bar u} _ 0}) / (v _ 0 - {{\bar v} _
0}), \:w _ 0 = {\bar w} _ 0 $.

For the solution of these equations we shall introduce the real
function $ \xi_1$ as the curvilinear integral:

$$
\xi _ 1 = \xi _ 1 (x, t) = i\int _ {(x _ 0, t _ 0)} ^ {(x, t)} (u
_ 0 - {\bar u} _ 0) dx + (v _ 0 - {\bar v} _ 0) dt, \:\:\:\:\xi _
1 = {\bar \xi} _ 1, \eqno (2.8)
$$
where $ x _ 0 $ is the initial point, $ t _ 0 $ is the initial
moment of time (later on we put $ t _ 0 = 0 $).

By virtue of a condition (2.5) the integral (2.8) does not depend
on the integration path. It allows to integrate the equation
(2.7). Taking into account that $ w _ 0 = \xi _ {1x} /\xi _ {1t}
$, we have:

$$ f _ 0 = \xi _ {1x} F _ 0 (\xi _ 1), \:\:\:\: g _ 0 = \xi _ {1x} G _ 0 (\xi _ 1), \eqno (2.9)
$$
where $ F _ 0 (.), \:G _ 0 (.) $ are arbitrary functions.
Therefore $ p _ 0 = \xi _ {1t} F _ 0 (\xi _ 1), \:q _ 0 = \xi _
{1t} G _ 0 (\xi _ 1) $, and finally we obtain:

$$
U = U (x, t) = \left (\begin {array} {cc} u_0 &  \xi_{1x} F _ 0
(\xi_ 1) \\ \xi _ {1x}  G _ 0 (\xi _ 1) &  {\bar u} _ 0 \end
{array} \right), \:\:\:V = V (x, t) = \left (\begin {array} {cc} v
_ 0 & \xi _ {1t} F _ 0 (\xi _ 1) \\ \xi _ {1t} G _ 0 (\xi _ 1)&
{\bar v} _ 0
\end {array} \right) .\eqno (2.10)
$$

Thus, using the Heisenberg picture, we formally have found the Lax
representation for equation (2.1). Let us notice that the
requirement of reality of potential together with the condition
(2.2) does not impose any additional restrictions on matrix
elements of matrices $ U $ and $ V $.

Let us consider in more detail the Heisenberg picture, which
appeared here. For that the equation (2.1) can be rewritten as:

$$
i\Psi ^ {(S)} _ t = H (t) \Psi ^ {(S)}, \eqno (2.11)
$$
where $ H (t) $ is the Hamiltonian, assumed hermitian and
time-dependent. Let's assume that

$$\Psi ^ {(S)} = {\hat U} (t) \Psi ^ {(H)}, \eqno (2.12)
$$
where $ \Psi ^ {(S)}, \:\Psi ^ {(H)} $ are the vectors of state in
the pictures of Shr\"odinger and Heisenberg, correspondingly, $
{\hat U} $ is the unitary operator. From (2.11), (2.12) it
follows, that this operator satisfies the equation: $ {\hat U} _ t
= -iH (t) {\hat U} $, whence by the assumption, that $ {\hat U}
(0) = I $, where $ I $ is the unit operator, we obtain:

$$
{ \hat U} (t) = {\hat T} e ^ {-i\int _ 0 ^ t H (t ^ {\prime}) dt ^
{\prime}}, \eqno (2.13)
$$
where $ {\hat T} $ is the time-ordering operator (Dayson
operator).

According to the basic principles of the Quantum Mechanics, if $ L
^ 0, \:L (t) $ are the observables in pictures of Shr\"odinger and
Heisenberg correspondingly, the average values of these
observables in both pictures coincide:

$$
< \Psi ^ {(S)} | L ^ 0 | \Psi ^ {(S)} > = < \Psi ^ {(H)} | L (t) |
\Psi ^ {(H)} > .
$$
Then taking into account the unitarity of $ {\hat U} $, from the
last equation we shall obtain

$$
L (t) = {\hat U} ^ {+} (t) L ^ 0 {\hat U} = {\hat U}^{-1} L ^ 0
{\hat U} . \eqno (2.14)
$$
where the symbol "+" \enskip means the hermitian conjugation.
Differentiating (2.14) in time, we shall have:

$$
L _ t = i\bigl [{\hat U} ^ {-1} H {\hat U}, \:L \bigr]. \eqno
(2.15)
$$
It is possible to obtain the representation equivalent to this,
using the another, in comparison with (2.13), relation for the
operator $ \hat U $. Actually, it is easy to see, that

$$
{ \hat U} (t) = 1-i\int _ 0 ^ tH (t ^ {\prime}) {\hat U} (t ^
{\prime}) dt ^ {\prime} =
$$
$$
= \sum _ {n = 0} ^ {\infty} \frac {1} {n!} \left (\frac {1} {i}
\right ) ^ n\int _ 0 ^ t dt _ 1\int _ 0 ^ t dt _ 2\ldots \int _ 0
^ t dt _ n {\hat T} (H (t _ 1) \ldots H (t _ n)). \eqno (2.16)
$$
Let us assume that $ {\hat R} = \bigl [H, \: {\hat U} \bigr] $.
Then due to (2.16) we have:

$$
{ \hat R} = \sum _ {n = 0} ^ {\infty} \frac {1} {n}! \left (\frac
{1} {i} \right ) ^ n\int _ 0 ^ t dt _ 1\int _ 0 ^ t dt _ 2\ldots
\int _ 0 ^ t dt _ n \bigl [H (t), \: {\hat T} (H (t _ 1) \ldots H
(t _ n)) \bigr],
$$
and, hence,

$$
L _ t = i\bigl [H + {\hat U} ^ {-1} {\hat R}, \:L\bigl] .\eqno
(2.17)
$$
Thus, equations (2.15) and (2.17) are the generalizations of
Heisenberg's equations of motion for the case when operators $ H
(t) $ and $ \hat U $ do not commute; and the equation (2.17) is
equivalent to (2.1). It is easy to show, supposing that the
quantity $ u _ 0 $ in (2.6) is the observable in a picture of
Heisenberg, setting formally $ u _ 0 = K _ 1\Psi ^ {(S)}, \:v _ 0
= K _ 2\Psi ^ {(S)} $, where $ K _ 1, \: K _ 2 $ are the integral
and integro-differential operators correspondingly, and obtaining
the closed equation for $ u _ 0 $: $ u _ {0t} = \bigl [K _ 2 (K _
1 ^ {-1} u _ 0) \bigr] _ x $.

\vskip0.7cm
\centerline {3. \bf {Darboux Transformation and Exact
Solutions.}}} \vskip0.4cm

Let us proceed to the construction of exact solutions of equation
(2.1). With that end in view, we shall use a method of matrix
Darboux Transformation [9], setting in (2.6) $ \alpha = 0 $.

Let $ \Phi _ 1 $ is the solution of system (2.3), corresponding to
a choice of solution $ \Psi = \Psi _ 1 $ of equation (2.1) with
some fixed value $ \lambda = \lambda _ 1 $. Let us assume that $ L
_ 1 = \Phi _ 1\Lambda _ 1 ^ {-1} \Phi _ 1 ^ {-1}, \:L _ 1 \in Mat
(2, \mathbb {C}) $. Then we shall check the covariance of the
system (2.3) under the transformation:

$$
\tilde \Phi = \Phi-L _ 1\Phi\Lambda.\eqno (3.1)
$$
This procedure results to the following "dressing"\enskip
formulas:

$$
\tilde U = L _ 1UL _ 1 ^ {-1}, \:\:\:\:\tilde U = U + \Phi _ 1
[\Lambda _ 1 ^ {-1}, \Phi _ 1 ^ {-1} \Phi _ {1x}] \Phi _ 1 ^ {-1},
$$
$$
\eqno (3.2)
$$
$$
\tilde V = L _ 1VL _ 1 ^ {-1}, \:\:\:\:\tilde V = V + \Phi _ 1
[\Lambda _ 1 ^ {-1}, \Phi _ 1 ^ {-1} \Phi _ {1t}] \Phi _ 1 ^ {-1},
$$
each of which, as it can be checked by direct calculations, is
equivalent to others.

The formulas (3.2) can be slightly simplified, using possible
symmetries (involutions) of the problem. Actually, combining the
relations $ \bar \Lambda = \sigma _ 1\Lambda\sigma _ 1 $ and $
\bar \Lambda = \sigma _ 2\Lambda \sigma _ 2 $, where $ \sigma _ 1,
\:\sigma _ 2 $ are Pauli matrices, and correspondingly: $ \bar U =
\sigma _ 1U\sigma _ 1 $ (if $ \bar p _ 0 = q _ 0 $) and $ \bar U =
\sigma _ 2U\sigma _ 2 $ (if $ \bar p _ 0 = -q _ 0 $), we obtain $
\bar \Phi = \sigma _ 1\Phi\sigma _ 1 $, and $ \bar \Phi = \sigma _
2\Phi\sigma _ 2 $, correspondingly.

From the second of relations (3.2) after some calculations we
have:

$$
\tilde \Psi = \Psi + iY _ x, \:\:Y = Y (x, \:t, \:\lambda _ 1)
\equiv \frac { \lambda _ 1 - {\bar \lambda _ 1}} {| \lambda _ 1 |
^ 2} \left (\frac {1} {1 - | y | ^ 2} \right), \eqno (3.3)
$$
where $ y = y (x, t, \lambda _ 1) = \varphi ^ {(1)} _ {21} (x, t,
\lambda _ 1) /\varphi ^ {(1)} _ {11} (x, t, \lambda _ 1) $, $
\varphi ^ {(1)} _ {ij} (x, t, \lambda _ 1) $ are matrix elements
of the matrix $ \Phi _ 1 $.

The expression for "dressed"\enskip potential of the NtSH equation
can be found either by replacing in (2.1) $ \Psi \to \tilde \Psi $
and $ W \to \tilde W $, or by using the following reasons. From
fourth of relations (3.2), analogously (3.3), we have:

$$
{ \tilde v} _ 0 = v _ 0-Y _ t. \eqno (3.4)
$$
From the explicit expression for $ v _ 0 $ we find:

$$
{ \tilde W} (x, t) = W (x, t) \frac {\Psi} {\Psi + iY} -\frac {Y _
{xt}} {\Psi + iY} -i\frac {Y _ {xx}} {\Psi + iY} .\eqno (3.5)
$$
Let us notice also that the reality of potential $ {\tilde W} $
follows from (3.3).

Now for construction of the solutions of equation (2.1) it is
necessary to integrate the system (2.3). For this purpose we shall
rewrite it as the overdetermined scalar system of the Riccatti
equations for the function $y$:

$$
y_x = -\lambda _ 1 [f _ 0y ^ 2 + (u _ 0 - {\bar u} _ 0) y - {\bar
f} _ 0], \:\:\: y _ t = -\lambda _ 1 [p _ 0y ^ 2 + (v _ 0 - {\bar
v} _ 0) y - {\bar p} _ 0] .\eqno (3.6)
$$
Therefore in particular it follows, that $ y $ satisfies the
linear equation in partial derivations: $ y _ x = w _ 0y _ t $,
which has the solution of a kind $ y = R (\xi _ 1) $, where $ R
(.) $ is an arbitrary complex-valued function of $ \xi _ 1 $,
determined by relation (2.8). To find this function, we shall
consider first of the equations (3.6) at $ t = 0 $ {\footnote {As
it follows from (2.8), the corresponding expression is $ \xi _ 1
(x, 0) = i\int _ {-M} ^ x (u _ 0 (x ^ {\prime}, 0) - {\bar u} _ 0
(x ^ {\prime}, 0)) dx $.}}. For its solutions we shall suppose,
that $ y = 0 $ at $ x = -M $ and $ y $ is a bounded function for
all $ x > M $. Let us take advantage of an arbitrariness in a
choice of function $ f _ 0 $. Then in the simplest case of
integrability it is necessary to choose $f_0 = - {\bar f} _ 0 = c
_ 0 = {\bar c} _ 0 = {\mathrm const}. $ Coming back to the
variable $ \xi _ 1 $ for the function $ y $, we shall find ( $
\lambda _ 1 = - {\bar \lambda _ 1}, \:y (x, \:t) = {\bar y} (x,
\:t)) $:

$$
Y (x, t) = y (\xi _ 1) = \frac {\tan \Theta _ 1} {\sqrt {2} + \tan
\Theta _ 1}, \:\:\Theta _ 1 = \Theta _ 1 (\xi _ 1) = -\frac
{\lambda _ 1} {\sqrt {2}} \int _ {-M} ^ {\xi _ 1} [u _ 0 (x ^
{\prime}, 0) - {\bar u} _ 0 (x ^ {\prime}, 0)] dx ^ {\prime}
.\eqno (3.7)
$$
Thus, from (3.3), taking into account (3.7), we shall have:

$$
\tilde {\Psi} = \Psi + 2i\frac {[u _ 0 (x, \:0) - {\bar U} _ 0 (x,
\:0)] [u _ 0 (x, \:t) - {\bar u} _ 0 (x, \:t)]} {\cosh ^ 2 \Theta
_ 1 (\sqrt {2} \tanh \Theta _ 1 + 1) ^ 2} (\sqrt {2} + \tanh
\Theta _ 1) .\eqno (3.8)
$$

The Darboux Transformation (3.1) can be multiply iterated, leading
multiply "dressed" \enskip solutions (procedure of their
construction for solitons of deformed Heisenberg magnet based on
the dressing formulas similar to (3.2) is given, for example, in
[10]).

Let us discuss two simplest examples, illustrating an application
of the proposed approach.

a). Let $ W (x, t) = 0 $, i.e. we shall consider the free NtSH
equation, and let us choose $ \Psi = \Psi _ 1 $ as a plane wave: $
\Psi _ 1 = ae ^ {ikx-ik ^ 2t} $, where $ a $ is the amplitude of
the wave, $ k $ is the wave number. Then, taking for simplicity $
M = 0 $, from (2.8) we have:

$$
\xi _ 1 = -\frac {4a} {k} \sin (kx-k ^ 2t), \eqno (3.9)
$$
and the expression for "dressed" \enskip wave function reads:

$$
\tilde \Psi = ae ^ {ikx-ik ^ 2t} \bigl [1-2a\frac {\cos kx ( 1-e ^
{-2i (kx-k ^ 2t)}) (\tanh \Theta _ 1 + \sqrt {2})} {\cosh ^ 2
\Theta _ 1 (\sqrt {2} \tanh \Theta _ 1 + 1) ^ 2} \bigr], \eqno
(3.10)
$$
where

$$
\Theta _ 1 = \Theta _ 1 (\xi _ 1) = \frac {i\lambda _ 1} {\sqrt
{2} k} a\sin \xi _ 1(x).
$$
This wave function will correspond to the potential $W (x, \:t)=
\tilde W (\xi _ 1) $, determined from (3.5). Let us notice here an
occurrence of the harmonic with the "opposite" \enskip phase in
the spectrum of excitations of the system.

b). Let the initial solution be the well-known wave function for
the harmonic oscillator ( $ W (x) = (1/4) \omega _ 0 ^ 2x ^ 2 $)
(see, for example, [11]):

$$\Psi _ {1n} (x, t) = c _ ne ^ {-i\omega _ 0 (n + 1/2) t} e ^ {-\frac {\omega _ 0x ^ 2} {4}}
H _ n (\px), \eqno (3.11)
$$
where $ \omega_0 $ is the frequency,  $ H_n (.) $ is Hermit
polynomial,  $ c _ n $ are the real normalization constants.
Taking $ \Psi _ {1n} (x, t) = X (x) T (t) $ and $ X = {\bar X} $,
according to (2.8), we shall find ($ M = \infty $):

$$
\xi _ 1 = - (T + {\bar T}) _ {| t = 0} \int _ {-\infty} ^ xXdx-iX
_ {x | x = -\infty} \int _ {0} ^ t (T - {\bar T}) dt + i\int _ 0 ^
t (T - {\bar T}) dt\int _ {-\infty} ^ xWXdx, \eqno (3.12)
$$
and

$$
\Theta _ 1 = \Theta _ 1 (\xi _ 1) = -ic _ n\lambda _ 1\int _ {-M}
^ {\xi _ 1} \ox H _ n (\px) dx, \:\:\:\lambda _ 1 = - {\bar
\lambda _ 1}.
$$
Then, from (3.3) we find

$$
\tilde \Psi _ {1n} (x, t) = c _ nH _ n (\px) e ^ {-\frac {1} {4}
\omega _ 0x ^ 2} \bigl [e ^ {-i\omega _ 0 (n + \frac {1} {2}) t} +
$$
$$
\eqno (3.13)
$$
$$ + \frac {2c _ nH _ n (\px) e ^ {-\frac {1} {4} \omega _ 0x ^ 2}
\cos [\omega _ 0 (n + \frac {1} {2}) t] (\sqrt {2} + \tanh \Theta
_ 1)} {\cosh ^ 2 \Theta_1 (\sqrt {2} \tanh \Theta _ 1 + 1) ^ 2}
\bigr],
$$
and, hence,  we have here the same effect, as in the previous
case.

In conclusion of this section we shall consider the connection of
Darboux Transformation and the theory of the coherent states. For
this purpose we shall represent the wave function of this state,
which is defined as the minimizing for the uncertainty relations
{\footnote {It is possible to define the coherent states in
another way as eigenstates of non-hermitian annihilation operator;
then the quantities $A_n$ are connected with matrix elements of
this operator.}}, as a decomposition in the wave functions of the
harmonic oscillator:

$$
\Psi _ {{\mathrm coher}} = \sum _ n A _ n \Psi _ {1n}, \eqno
(3.14)
$$
where $ \Psi _ {1n} $ are defined by the expression (3.11), and $
A _ n $ are coefficients of the decomposition. After the single
application of Darboux Transformation we have: $ A _ n \to {\tilde
A} _ n, \:\Psi _ {1n} \to {\tilde \Psi} _ {1n} $ and, besides
that,

$$
\sum _ n A _ n \Psi _ {1n} = \sum _ n {\tilde A} _ n {\tilde \Psi}
_ {1n}, \eqno (3.15)
$$
where

$$
A _ n = \int \Psi _ {\mathrm coher} (x, t) {\bar {\Psi} _ {1n}}
(x, t) \:dx, \:\: {\tilde A} _ n = \int \Psi _ {\mathrm coher} (x,
t) {\bar {{\tilde \Psi}} _ {1n}} (x, t) \:dx.
$$
Then taking into account (3.3), we obtain the connection of
coefficients $ A_n$ and $ {\tilde A}_n$:

$$
{ \tilde A} _ n = \tilde A _ n-i\int \Psi _ {\mathrm {coher}}
{\bar Y} _ x\:dx, \eqno (3.16)
$$
where the quantity $ Y $ is determined in (3.3).

\vskip1cm \vskip1cm \hfil {4}. {\bf Conclusion.}\hfil \vskip0.3cm

Let us notice, that it is possible also to find a slightly
different Lax representation for the equation (2.1), allowing to
study the scattering theory for the operator of the associated
linear problem. On this basis it is possible to obtain, in
particular, "the relations of balance" \enskip (instead of
conservation laws, because an initial system, as it was already
marked, is not conservative), and also to construct the hierarchy
of the "highest" equations. It is especially interesting, that
already the simplest equations appear to be nonlinear.

As a whole, such approach can probably serve as a basis for
quasiclassical quantization of NtSH equation as a completely
integrable system. We are going to present these results later on.

The author is grateful to I.V.Komarov and M.V. Ioffe for the
useful critical remarks.

\vskip 2cm

\vskip0.3cm 1. \parbox [t] {12.7cm} {{\em B.G.Bagrov,
B.F.Samsonov.} Elementarnye chasticy i atomnoe jadro {\bf 28}, 951
(1997).}

\vskip0.3cm 2. \parbox [t] {12.7cm} {{\em B.F.Samsonov.} JETP {\bf
114}, 1930 (1998).}

\vskip0.3cm 3. \parbox [t] {12.7cm} {{\em I.A.Dynnikov,
S.P.Novikov.} Uspekhi mathemathicheskich nauk {\bf 52}, 175
(1997).}

\vskip0.3cm 4.
\parbox [t] {12.7cm} {{\em B.G.Machan'kov, Y.P.Rybakov, V.I.Sanuk.}
УФН {\bf 164}, 121 (1994).}

\vskip0.3cm 5. \parbox [t] {12.7cm} {{\em S.V.Manakov.} Physica D.
{\bf 3}, 420 (1981).}

 \vskip0.3cm 6.
\parbox [t] {12.7cm} {{\em E.Sh. Gutshabash, M.A.Salle.} In:
Proceedings of International Seminar "Day on Diffraction-2001",
pp.116-127, St-Petersburg, (2001).}

\vskip0.3cm 7. \parbox [t] {12.7cm} {{\em A.I.Baz',
Ya.B.Zeldovitch, A.M. Perelomov.} Scattering, reactions and
decompositions in nonrelativistic quantum mechanics. М., Nauka,
(1971).}

\vskip0.3cm 8.
\parbox [t] {12.7cm} {{\em A.N.Leznov, V.I.Man'ko, M.A.Saveliev.}
In "Solitons, instantons and operator quantization". Proceedings
of PhiAN, т.165, М., Nauka, (1986).}

\vskip0.3cm 9.
\parbox [t] {12.7cm}
{ {\em M.А. Salle, V.B.Matveev. Darboux Transformation and
Solitons. Springer-Verlag (1991).}}

\vskip0.3cm \vskip0.3cm 10. \parbox [t] {12.cm} {{\em
E.Sh.Gutshabash.} JETP Letters {\bf 73}, 317 (2001).}

\vskip0.3cm \vskip0.3cm 11. \parbox [t] {12.cm} {{\em S.Flugge.}
Practical Quantum mechanics. V.1. Berlin-Heidelberg-New-York.,
Springer-Verlag (1971).}

\end {document}